\documentclass[twocolumn,american,showpacs,prl]{revtex4-1}
\usepackage[T1]{fontenc}
\usepackage[latin9]{inputenc}
\usepackage{color}
\usepackage{babel}
\usepackage{amsmath}
\usepackage{amssymb}
\usepackage{graphicx}
\usepackage{esint}
\usepackage[unicode=true,pdfusetitle,
 bookmarks=true,bookmarksnumbered=false,bookmarksopen=false,
 breaklinks=false,pdfborder={0 0 0},backref=false,colorlinks=true]
 {hyperref}
\usepackage{breakurl}

\makeatletter
 
 \@ifundefined{textcolor}{}
 {%
   \definecolor{BLACK}{gray}{0}
   \definecolor{WHITE}{gray}{1}
   \definecolor{RED}{rgb}{1,0,0}
   \definecolor{GREEN}{rgb}{0,1,0}
   \definecolor{BLUE}{rgb}{0,0,1}
   \definecolor{CYAN}{cmyk}{1,0,0,0}
   \definecolor{MAGENTA}{cmyk}{0,1,0,0}
   \definecolor{YELLOW}{cmyk}{0,0,1,0}
 }

\usepackage{babel}
\usepackage{babel}

\makeatother

\begin{document}

\title{Controlling the position of traveling waves in reaction-diffusion
systems}

\author{Jakob Löber}

\email{jakob@physik.tu-berlin.de}

\author{Harald Engel}

\address{Institut für Theoretische Physik, EW 7-1, Technische Universität
Berlin, Hardenbergstraße 36, 10623 Berlin, Germany}


\begin{abstract}
We present a method to control the position as a function of time
of one-dimensional traveling wave solutions to reaction-diffusion
systems according to a pre-specified protocol of motion. Given this
protocol, the control function is found as the solution of a perturbatively
derived integral equation. Two cases are considered. First, we derive
an analytical expression for the space ($x$) and time ($t$) dependent
control function $f\left(x,t\right)$ that is valid for arbitrary
protocols and many reaction-diffusion systems. These results are close
to numerically computed optimal controls. Second, for stationary control
of traveling waves in one-component systems, the integral equation
reduces to a Fredholm integral equation of the first kind. In both
cases, the control can be expressed in terms of the uncontrolled wave
profile and its propagation velocity, rendering detailed knowledge
of the reaction kinetics unnecessary. 
\end{abstract}
\maketitle
\begin{figure}
\includegraphics{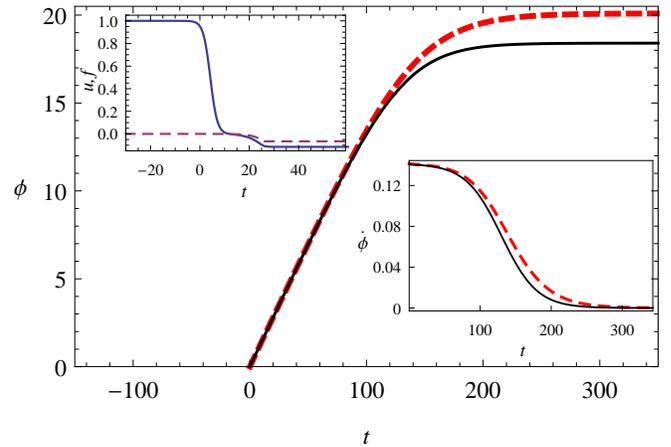} \caption{\label{fig:StationaryControlSchloegl}Deceleration of a Schlögl front
by an additive stationary control. Red dashed line is the result of
numerical simulations, black solid line is the pre-given protocol.
Shown are the position $\phi$ and the velocity $\dot{\phi}$ (right
inset). The front profile (blue solid line) is slightly deformed in
the region where the control (purple dotted line) is large, see left
inset. Compare also S8 in \cite{supplement}.}
\end{figure}

A variety of approaches have been developed for the purposeful manipulation
of reaction-diffusion (RD) systems as e.g. the application of feedback-mediated
control loops with and without delays, external spatio-temporal forcing
or imposing heterogeneities and geometric constraints on the medium
\cite{mikhailov2006control,*vanag2008design}. For example, unstable
patterns can be stabilized by global feedback control, as was shown
in experiments with the light-sensitive Belousov-Zhabotinsky (BZ)
reaction \cite{mihaliuk2002feedback,*Zykov2004global,*zykov2004feedback}.
Two feedback loops were used to stabilize unstable wave segments and
to guide their propagation direction \cite{sakurai2002design}. Position
control, or dragging, of a traveling chemical pulse \cite{wolff2003gentle}
on an addressable catalyst surface \cite{wolff2001spatiotemporal}
was accomplished experimentally by a moving, localized temperature
heterogeneity. Dragging of fronts in chemical and phase transitions
models as well as targeted transfer of nonlinear Schrödinger pulses
by moving heterogeneities was studied in \cite{kevrekidis2004dragging,*nistazakis2002targeted,*malomed2002pulled}.\\
Many of these control methods rely on extensive knowledge about the
system to be controlled. Feedback control necessitates continuous
monitoring of the system, while optimal control \cite{troltzsch2010optimal,*jorge1999numerical,*theissen2006optimale,*buchholz2013on}
requires full knowledge of the underlying partial differential equations
(PDE) governing the system's evolution in time and space.\\
In this Letter we propose a method which partially overcomes the aforementioned
difficulties and still compares favorably with a competing control
method, namely optimal control. We consider the problem to control
the position over time of one-dimensional traveling waves (TW) by
spatio-temporal forcing. The starting point is a system of RD equations
\begin{align}
\partial_{t}\mathbf{u} & =D\partial_{x}^{2}\mathbf{u}+\mathbf{R}\left(\mathbf{u}\right)+\epsilon\mathcal{G}\left(\mathbf{u}\right)\mathbf{f}\left(x,t\right),\label{eq:ReactionDiffusionSystem}
\end{align}
where $D$ is a diagonal matrix of constant diffusion coefficients,
$\mathbf{f}$ is a spatio-temporal perturbation, $\mathcal{G}$ a
(possibly $\mathbf{u}$-dependent) coupling matrix, and $\mathbf{R}$
the nonlinear reaction kinetics. The unperturbed ($\epsilon=0$) solution
$\mathbf{U}_{c}\left(\xi\right)$, $\xi=x-ct$, is assumed to be a
TW, stationary in the reference frame co-moving with velocity $c$,
so that 
\begin{align}
D\partial_{\xi}^{2}\mathbf{U}_{c}\left(\xi\right)+c\partial_{\xi}\mathbf{U}_{c}\left(\xi\right)+\mathbf{R}\left(\mathbf{U}_{c}\left(\xi\right)\right) & =0.\label{eq:TravellingWave}
\end{align}
The eigenvalues of the linear operator 
\begin{align}
\mathcal{L} & =D\partial_{\xi}^{2}+c\partial_{\xi}+\mathcal{D}\mathbf{R}\left(\mathbf{U}_{c}\left(\xi\right)\right)\label{eq:LinearOperatorL}
\end{align}
determine the stability of the TW, where $\mathcal{D}\mathbf{R}\left(\mathbf{U}_{c}\left(\xi\right)\right)$
denotes the Jacobian matrix of $\mathbf{R}$ evaluated at the TW.
We assume $\mathbf{U}_{c}$ to be stable. Therefore the eigenvalue
of $\mathcal{L}$ with largest real part is $\lambda_{0}=0$, and
the Goldstone mode $\mathbf{W}\left(\xi\right)=\mathbf{U}_{c}'\left(\xi\right)$,
also called propagator mode, is the corresponding eigenfunction. Because
$\mathcal{L}$ is in general not self-adjoint, the eigenfunction $\mathbf{W}^{\dagger}\left(\xi\right)$
of the adjoint operator $\mathcal{L}^{\dagger}$ to eigenvalue zero,
the so-called response function, is not identical to $\mathbf{W}\left(\xi\right)$.
Expanding Eq. \eqref{eq:ReactionDiffusionSystem} with $\mathbf{u}=\mathbf{U}_{c}+\epsilon\mathbf{v}$
up to $\mathcal{O}\left(\epsilon\right)$ yields a PDE $\partial_{t}\mathbf{v}=\mathcal{L}\mathbf{v}+\mathcal{G}\mathbf{f}$.
Its solution $\mathbf{v}$ can be expressed in terms of eigenfunctions
$\mathbf{w}_{i}$ of $\mathcal{L}$ as ${\mathbf{v}\left(\xi,t\right)=\sum_{i}a_{i}\left(t\right)\mathbf{w}_{i}\left(\xi\right)}$
with expansion coefficients ${a_{i}\sim\intop_{t_{0}}^{t}d\tilde{t}e^{\lambda_{i}\left(t-\tilde{t}\right)}b\left(\tilde{t}\right)}$
and $b$ a functional of $\mathbf{f}$ involving eigenfunctions of
$\mathcal{L}^{\dagger}$ \cite{supplement}.\\
By multiple scale perturbation theory for small $\epsilon$, the following
equation of motion (EOM) for the position $\phi\left(t\right)$ of
the TW in the presence of the spatio-temporal perturbation $\mathbf{f}$
can be obtained, 
\begin{align}
\dot{\phi} & =c-\frac{\epsilon}{K_{c}}\int_{-\infty}^{\infty}dx\mathbf{W}^{\dagger T}\left(x\right)\mathcal{G}\left(\mathbf{U}_{c}\left(x\right)\right)\mathbf{f}\left(x+\phi,t\right),\label{eq:EquationOfMotion}
\end{align}
with constant $K_{c}=\intop_{-\infty}^{\infty}dx\mathbf{W}^{\dagger T}\left(x\right)\mathbf{U}_{c}'\left(x\right)$
and initial condition $\phi\left(t_{0}\right)=\phi_{0}$. For monotonously
decreasing front solutions, we define its position as the point of
steepest slope, while for pulse solutions it is the point of maximum
amplitude of an arbitrary component.\\
The EOM Eq. \eqref{eq:EquationOfMotion} only takes into account the
contribution of the perturbation $\mathbf{f}$ which affects the position
of the TW. Adding to the TW a small term proportional to the Goldstone
mode slightly shifts the TW because (for details compare \cite{supplement})
\begin{align}
\mathbf{U}_{c}\left(x-ct\right)+\epsilon p\partial_{x}\mathbf{U}_{c}\left(x-ct\right) & \approx\mathbf{U}_{c}\left(x-ct+\epsilon p\right).\label{eq:GoldstoneModeEffect-1}
\end{align}
Due to the orthogonality of eigenmodes $\mathbf{w}_{i}$ to different
eigenvalues $\lambda_{i}$, the Goldstone mode alone accounts for
propagation, while all other modes account for the deformation of
the wave profile $\mathbf{U}_{c}$. The spectral gap $d>0$, i.e.
the separation between $\lambda_{0}=0$ and the real part of the next
largest eigenvalue, characterizes the deformation relaxation time
scale. The larger $d$ the faster decay all deformation modes for
large times as long as the perturbation $\mathbf{f}$ remains bounded
in time. Secular growth of the expansion coefficient $a_{0}$ arising
even for bounded perturbations is prevented by assuming that $p$
depends on a slow time scale $T=\epsilon t$ and applying a solvability
condition. The EOM Eq. \eqref{eq:EquationOfMotion} must be seen as
the first two terms of an asymptotic series with bookkeeping parameter
$\epsilon$ \cite{bender1978advanced}. In the following we set $\epsilon=1$
and expect Eq. \eqref{eq:EquationOfMotion} to be accurate only if
the perturbation $\mathbf{f}$ is sufficiently small in amplitude.
For a detailed derivation and applications of Eq. \eqref{eq:EquationOfMotion}
compare \cite{loeber2012front} and \cite{schimanskygeier1983effect,*engel1985noise,*engel1987interaction,*kulka1995influence,*bode1997front}.
Methods closely related to the derivation of EOM Eq. \eqref{eq:EquationOfMotion}
are e.g. phase reduction methods for limit cycle solutions to dynamical
systems \cite{pikovsky2003synchronization} and the soliton perturbation
theory \cite{yang2011nonlinear} developed for nonlinear conservative
systems supporting TWs as e.g. the Korteweg-de Vries equation.\\
In this Letter, we do not perceive Eq. \eqref{eq:EquationOfMotion}
as an ordinary differential equation for the position $\phi\left(t\right)$
of the wave under the given perturbation $\mathbf{f}$. Instead, Eq.
\eqref{eq:EquationOfMotion} is viewed as an integral equation for
the \textit{control function} $\mathbf{f}$. The idea is to find a
control which solely drives propagation in space according to an arbitrary
given \textit{protocol of motion} $\phi\left(t\right)$. Simultaneously,
we expect $\mathbf{f}$ to prevent large deformations of the uncontrolled
wave profile $\mathbf{U}_{c}\left(\xi\right)$. Expressed in the language
of eigenmodes of $\mathcal{L}$, we search for a control $\mathbf{f}$
which excites the Goldstone mode $\mathbf{U}_{c}'\left(\xi\right)$
in an appropriate manner and minimizes excitation of all modes responsible
for the deformation of the wave profile. We assume that the wave moves
unperturbed until reaching position $\phi_{0}$ at time $t_{0}$,
upon which the control is switched on.\\
A general solution of the integral equation Eq. \eqref{eq:EquationOfMotion}
for the control $\mathbf{f}$ with given protocol of motion $\phi\left(t\right)$
is 
\begin{align}
\mathbf{f}\left(x,t\right) & =\left(c-\dot{\phi}\right)\dfrac{K_{c}}{G_{c}}\,\mathcal{G}^{-1}\left(\mathbf{U}_{c}\left(x-\phi\right)\right)\mathbf{h}\left(x-\phi\right),\label{eq:GeneralControl}
\end{align}
with constant $G_{c}=\intop_{-\infty}^{\infty}dx\mathbf{W}^{\dagger T}\left(x\right)\mathbf{h}\left(x\right)$.
Here $\mathcal{G}^{-1}$ denotes the matrix inverse to $\mathcal{G}$.
The profile $\mathcal{G}^{-1}\mathbf{h}$ of the control $\mathbf{f}$
is co-moving with the controlled wave while the time dependent coefficient
$c-\dot{\phi}$ determines the control amplitude. Eq. \eqref{eq:GeneralControl}
contains a so far undefined arbitrary function $\mathbf{h}\left(x\right)$.
A control proportional to the Goldstone mode $\mathbf{U}_{c}'$ shifts
the TW as a whole, simultaneously preventing large deformations of
the wave profile \cite{supplement}. Therefore, in the following we
choose $\mathbf{h}\left(x\right)=\mathbf{U}_{c}'\left(x\right)$,
i.e. 
\begin{align}
\mathbf{f}\left(x,t\right) & =\left(c-\dot{\phi}\right)\mathcal{G}^{-1}\left(\mathbf{U}_{c}\left(x-\phi\right)\right)\mathbf{U}_{c}'\left(x-\phi\right).\label{eq:SpatiotemporalControl}
\end{align}
Because $K_{c}=G_{c}$ in this case, the solution does not contain
the response function $\mathbf{W}^{\dagger}$.\\
In the examples discussed below, the given protocol $\phi\left(t\right)$
is compared with position over time data obtained by numerical simulations
of the controlled RDS subjected to no-flux or periodic boundary conditions
and $\mathbf{U}_{c}\left(x-\phi_{0}\right)$ as the initial condition.
Furthermore, the result Eq. \eqref{eq:SpatiotemporalControl} is compared
with optimal control solutions obtained by numerically minimizing
the constrained functional on the spatio-temporal domain $Q$ \cite{troltzsch2010optimal,*jorge1999numerical,*theissen2006optimale,*buchholz2013on}
\begin{align}
\mathcal{J} & =\dfrac{1}{2}\iint_{Q}dxdt\left|\left|\mathbf{u}-\mathbf{u}_{d}\right|\right|^{2}+\dfrac{\lambda}{2}\iint_{Q}dxdt\left|\left|\mathbf{f}\right|\right|^{2}.
\end{align}
Here, $\lambda$ is a small ($\approx10^{-6}$) regularization parameter
and $\mathbf{u}$ is constrained to be a solution of the controlled
RDS Eq. \eqref{eq:ReactionDiffusionSystem}. $\mathbf{u}_{d}$ denotes
an arbitrary desired spatio-temporal distribution which we want to
enforce onto the system. For the purpose of position control, $\mathbf{u}_{d}$
is a TW shifted according to the protocol $\phi$,
\begin{figure}
\includegraphics{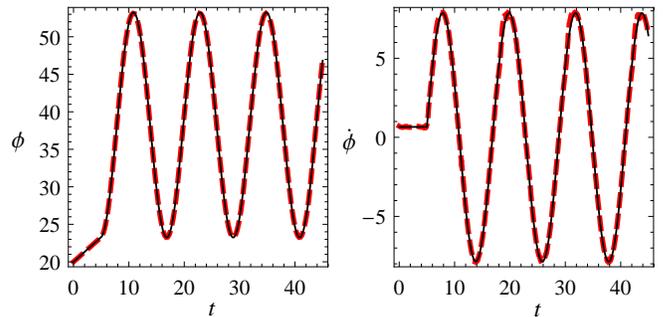} \caption{\label{fig:SchloeglControl}Periodic acceleration and deceleration
of a Schlögl front realized by multiplicative control. Left: The numerically
obtained front position (red dashed line) is in excellent agreement
with the protocol of motion $\phi\left(t\right)=B_{0}+A\sin\left(2\pi t/T+B_{1}\right)$
(black solid line). $B_{0/1}$ are determined by $\phi\left(t_{0}\right)=\phi_{0},\,\dot{\phi}\left(t_{0}\right)=c$
so that the protocol is smooth at $t=t_{0}$. Right figure shows velocities.
See S1 in \cite{supplement} for a movie.}
\end{figure}
 
\begin{align}
\mathbf{u}_{d}\left(x,t\right) & =\mathbf{U}_{c}\left(x-\phi\left(t\right)\right).
\end{align}
The coupling matrix $\mathcal{G}$ depends upon the ability to control
system parameters in a spatio-temporal way. In general, if $\mathbf{R}\left(\mathbf{u};\mathbf{p}\right)$
depends on the controllable parameters $\mathbf{p}$, we substitute
$\mathbf{p}\rightarrow\mathbf{p}+\epsilon\mathbf{f}$, expand in $\epsilon$,
and define the coupling matrix by ${\mathcal{G}\left(\mathbf{u}\right)=\partial\mathbf{R}\left(\mathbf{u};\mathbf{p}\right)/\partial\mathbf{p}}$.
As an example, we consider an autocatalytic chemical reaction mechanism
proposed by Schlögl ${A_{1}+2X\overset{k_{1}^{+}}{\underset{k_{1}^{-}}{\rightleftharpoons}}3X,\, X\overset{k_{2}^{+}}{\underset{k_{2}^{-}}{\rightleftharpoons}}A_{2}}$
\cite{Schlogl1972crm}. Under the assumption that the concentrations
$c_{1/2}=\left[A_{1/2}\right]$ of the chemical species $A_{1/2}$
are kept constant in space and time, a cubic reaction function ${R\left(u\right)=k_{1}^{+}c_{1}u{}^{2}-k_{1}^{-}u{}^{3}-k_{2}^{+}u+k_{2}^{-}c_{2}}$
dictates the time evolution of the concentration $u=\left[X\right]$.
We assume that the concentrations $c_{1/2}$ can be controlled spatio-temporally,
i.e., $c_{1/2}\rightarrow c_{1/2}+\epsilon f\left(x,t\right)$. Control
by $c_{2}$ will be additive with $\mathcal{G}\left(u\right)=k_{2}^{-}$,
while for control via $c_{1}$ the spatio-temporal forcing couples
multiplicatively to the RD kinetics and $\mathcal{G}\left(u\right)=k_{1}^{+}u^{2}$.
A different example for position control, realized experimentally
in \cite{wolff2003gentle}, exploits the dependency of the rate coefficients
$k_{1/2}^{\pm}$ on temperature $T$ according to the Arrhenius law
$k\sim e^{-E/\left(k_{B}T\right)}$. Substituting $T\rightarrow T+\epsilon f\left(x,t\right)$
and expansion in $\epsilon$ yields the coupling function $\mathcal{G}\left(u\right)$.
\begin{figure}
\includegraphics{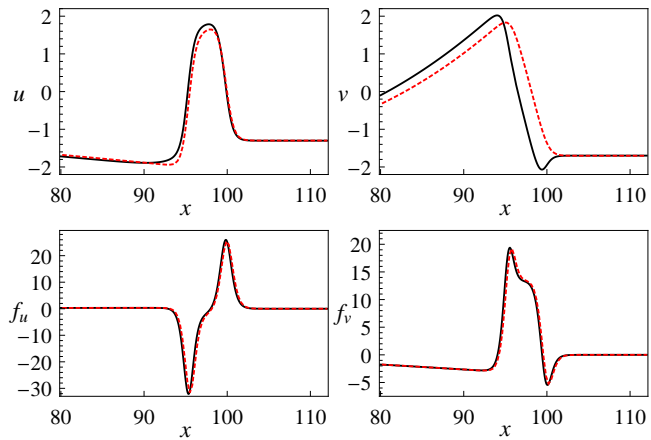}

\caption{\label{fig:FHNControl}Snapshot of position control of a FHN pulse
with an invertible coupling matrix $\mathcal{G}$ taken from movie
S2 in \cite{supplement}. Results by analytical control (black solid)
agree very well with results obtained by optimal control (red dashed).
Clockwise from top left: activator $u$, inhibitor $v$, controls
$f{}_{u},\, f_{v}$.}
\end{figure}
In the bistable parameter regime, the unperturbed Schlögl model has
an analytically known traveling front solution $U_{c}$ connecting
the stable and the metastable homogeneous steady state as $x\rightarrow\pm\infty$
\cite{Schlogl1972crm}. Suppose we want to move the front periodically
back and forth in a sinusoidal manner via a spatio-temporal control
of parameter $c_{1}$. Fig. \ref{fig:SchloeglControl} left shows
that the numerically obtained front position follows the protocol
very closely. The maximum enforced front velocity, $\max_{t}\dot{\phi}\left(t\right)=7.854$,
is much larger than the velocity $c=0.662$ of the uncontrolled front,
compare Fig. \ref{fig:SchloeglControl} right.
\begin{figure}
\includegraphics{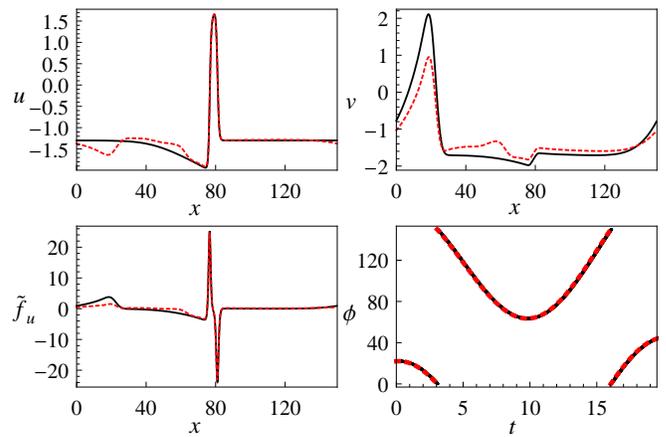}

\caption{\label{fig:FHNActivatorControl}Snapshot of position control of a
FHN pulse with a non-invertible coupling matrix $\mathcal{G}$. The
control $\tilde{f}_{u}$ (bottom left) acts solely on the activator
equation. The controlled inhibitor pulse profile (top right) is much
more deformed than the activator pulse profile (top left). Shown are
results of optimal (red dashed) and analytical control (black solid).
Bottom right: Analytical protocol (black solid) and numerically obtained
position over time data for the maximum activator value of the controlled
RDS (red dashed). See movie S3 \cite{supplement}.}
\end{figure}
\\
Now we apply position control to the stable traveling pulse solution
of FitzHugh-Nagumo (FHN) equations

\begin{align}
\partial_{t}u= & D_{u}\partial_{x}^{2}u+f_{1}\left(u\right)-v+\epsilon\left(\mathcal{G}_{11}f_{u}+\mathcal{G}_{12}f_{v}\right),\label{eq:FitzHughNagumo}\\
\partial_{t}v= & D_{v}\partial_{x}^{2}v+\tilde{\epsilon}\left(u-\delta\right)-\tilde{\epsilon}\gamma v+\epsilon\left(\mathcal{G}_{21}f_{u}+\mathcal{G}_{22}f_{v}\right),\nonumber 
\end{align}
where ${f_{1}=3u-u^{3}}$ and $\mathcal{G}_{ij}$ denote the components
of the coupling matrix $\mathcal{G}$. As an example, we consider
an accelerating protocol $\phi\left(t\right)=ct\left(1+t/4\right)$.
We assume that two additive parameters can be controlled independently.
For the choice ${\mathcal{G}=\small\left(\begin{array}{cc}
1 & 0\\
1/2 & 1
\end{array}\right)}$, $\mathcal{G}$ is invertible. The obtained control function as well
as the controlled pulse profile are close to the corresponding results
obtained by an optimal control, see Fig. \ref{fig:FHNControl}.\\
If the coupling matrix $\mathcal{G}$ is not invertible, Eq. \eqref{eq:SpatiotemporalControl}
for the control cannot be used. Because the inhibitor kinetics is
linear in $v$, Eq. \eqref{eq:FitzHughNagumo} can be written as a
single nonlinear integro-differential equation (IDE) for the activator
$u$ 
\begin{align}
\partial_{t}u & =D_{u}\partial_{x}^{2}u+f_{1}\left(u\right)-\mathcal{K}\left(\tilde{\epsilon}\left(u-\delta\right)+\epsilon f_{v}\right)-\mathcal{K}_{0}v_{0}+\epsilon f_{u}.\label{eq:IntegroDiff}
\end{align}
$\mathcal{K}$ and $\mathcal{K}_{0}$ are integral operators, involving
Green's function, of the inhomogeneous linear PDE for the inhibitor
$v$ with initial condition $v\left(x,t_{0}\right)=v_{0}\left(x\right)$
\begin{align}
\partial_{t}v-D_{v}\partial_{x}^{2}v+\tilde{\epsilon}\gamma v & =\tilde{\epsilon}\left(u-\delta\right)+\epsilon f_{v}.\label{eq:EqInhibitor}
\end{align}
We contrast Eq. \eqref{eq:IntegroDiff} with the equation obtained
from Eq. \eqref{eq:IntegroDiff} by substituting $f_{u}\rightarrow\tilde{f}_{u},\, f_{v}\rightarrow0$.
Comparing the control terms yields the control $\tilde{f}_{u}$ acting
solely on the activator equation, 
\begin{align}
\tilde{f}_{u} & =-\mathcal{K}f_{v}+f_{u},\label{eq:activatorcontrol}
\end{align}
were $f_{u}$ and $f_{v}$ are given by Eq. \eqref{eq:SpatiotemporalControl}
with $\mathcal{G}\equiv1$. We apply the control $\tilde{f}_{u}$
with a sinusoidal protocol to a FHN pulse. The activator's maximum
follows the protocol closely, see bottom right of Fig. \ref{fig:FHNActivatorControl}.
Comparing the result for $\tilde{f}_{u}$, Eq. \eqref{eq:activatorcontrol},
with an optimal control result reveals good overall agreement (bottom
left of Fig. \ref{fig:FHNActivatorControl}). However, for both control
methods the inhibitor profile (top right) is largely deformed although
the activator profile remains comparably unaffected (top left). Reduction
of the RD equations to a single IDE and thereby derivation of a control
is possible for, but not restricted to, all models of the form \cite{supplement}
\begin{align}
\partial_{t}u & =D_{u}\partial_{x}^{2}u+f\left(u,v_{2},\dots,v_{n}\right)+\epsilon f_{u},\label{eq:HodgkinHuxleyType}\\
\partial_{t}v_{i} & =D_{i}\partial_{x}^{2}v_{i}+h_{i}\left(u\right)v_{i}+g_{i}\left(u\right)+\epsilon f_{i},\, i\in\left\{ 2,\dots,n\right\} .\nonumber 
\end{align}
This class includes Hodgkin-Huxley type models (with ${D_{i}=0}$)
for the action potential propagation in neuronal and cardiac tissue
\cite{murray1993mathematical}. The modified Oregonator model describing
the light-sensitive BZ reaction \cite{krug1990analysis} is not of
the form Eq. \eqref{eq:HodgkinHuxleyType} but can nevertheless be
written as a single IDE. We present position control of chemical concentration
waves in the photosensitive BZ reaction applying actinic light of
space-time dependent intensity to the reaction in the supplemental
material S6 \cite{supplement}.\\
In many experiments, a stationary control $f\left(x\right)$ is much
less demanding to realize than a spatio-temporal control $f\left(x,t\right)$.
For single component RD systems, we can formulate a Fredholm integral
equation of the first kind for $f\left(x\right)$ 
\begin{align}
g\left(\phi\right)=cK_{c}-\frac{K_{c}}{T'\left(\phi\right)} & =\int_{-\infty}^{\infty}dxK\left(\phi-x\right)f\left(x\right),\label{eq:FredholmIntegralEquation}
\end{align}
with kernel ${K\left(x\right)=e^{-cx/D}U_{c}'\left(-x\right)\mathcal{G}\left(U_{c}\left(-x\right)\right)}$
and inhomogeneity $g$. We introduced the inverse function ${T=\phi^{-1}}$
and used the general expression for the adjoint Goldstone mode for
single component systems, ${W^{\dagger}\left(x\right)=e^{cx/D}U_{c}'\left(x\right)}$.
Eq. \eqref{eq:FredholmIntegralEquation} can be solved with the help
of the convolution theorem for the two-sided Laplace transform, see
\cite{supplement}.\\
As an example, we choose a protocol which drives the propagation velocity
to zero according to 
\begin{align}
\dot{\phi}\left(t\right) & =\dfrac{c}{2}\left(1+\tanh\left(k\left(t_{1}-t\right)\right)\right),\; t_{1}>t_{0},\, k>0.\label{eq:VelocityProtocol}
\end{align}
In the limit $k\rightarrow\infty$, this protocol would stop the front
instantaneously at time $t=t_{1}$ because $\lim_{k\rightarrow\infty}\dot{\phi}\left(t\right)=c\Theta\left(t_{1}-t\right)$,
where $\Theta$ represents the Heaviside Theta function. For the inhomogeneity
$g$ we find 
\begin{align}
g\left(\phi\right) & =K_{c}\frac{c\exp\left(\frac{2k}{c}\left(ct_{0}+\phi-\phi_{0}\right)\right)}{e^{2kt_{0}}+e^{2kt_{1}}}.
\end{align}
An additive control with $\mathcal{G}\left(u\right)=1$ is assumed.\\
We consider a rescaled Schlögl model with reaction function $R\left(u\right)=-u\left(u-a\right)\left(u-1\right)$.
The front solution is given as $U_{c}\left(\xi\right)=1/\left(1+\exp\left(\xi/\sqrt{2}\right)\right)$
with propagation velocity $c=\left(1-2a\right)/\sqrt{2}$ for $D=1$.
The region of convergence of the Laplace transforms of kernel $K$
and inhomogeneity $g$ determines the range of allowed values for
$k$ as $0<k<c\left(1/\sqrt{2}-c\right)/2=k_{\text{max}}$. This amounts
to a minimum acceleration (or maximum deceleration) at time $t=t_{1}$
equal to 
\begin{align}
\ddot{\phi}\left(t_{1}\right)= & -\dfrac{c}{2}k>-\left(\dfrac{c}{2}\right)^{2}\left(\dfrac{1}{\sqrt{2}}-c\right),
\end{align}
 which can be realized under this control given explicitly by 
\begin{align}
f\left(x\right) & =-\frac{K_{c}c^{2}\sin\left(\frac{\sqrt{2}\pi}{c}\left(c^{2}+2k\right)\right)e^{\frac{2k}{c}\left(x-\phi_{0}\right)}}{\sqrt{2}\pi\left(1+e^{2k\left(t_{1}-t_{0}\right)}\right)\left(c^{2}+2k\right)}.\label{eq:ControlSolution}
\end{align}
The divergence for $x\rightarrow\infty$ can be circumvented by cutting
off $f$ in such a way that ${R\left(u\right)+\epsilon f\left(x\right)=0}$
locally keeps three different real roots, meaning that bistability
is preserved at every point in space. A more systematic approach to
prevent divergence of $f\left(x\right)$ would be to consider the
Fredholm integral equation Eq. \eqref{eq:FredholmIntegralEquation}
supplemented with inequality constraints ${f_{\text{min}}\leq f\leq f_{\text{max}}}$
for the control function.\\
Under the control Eq. \eqref{eq:ControlSolution}, the velocity of
the numerical solution first follows the protocol velocity closely,
see right inset of Fig. \ref{fig:StationaryControlSchloegl}. Deviations
arise when the transition region of the front enters the domain with
large absolute values of the control. These velocity deviations accumulate
to a difference in the position at which the front is stopped. The
front profile is slightly deformed in the region where the control
is large because the solution Eq. \eqref{eq:ControlSolution} is not
proportional to the Goldstone mode, see left inset in Fig. \ref{fig:StationaryControlSchloegl}.
\\
In conclusion, we have demonstrated that the proposed method is well-suited
to control the position of traveling fronts and pulses in RD systems
according to a pre-given protocol of motion $\phi\left(t\right)$
while preserving the profile $\mathbf{U}_{c}$ of the uncontrolled
wave. To determine the control functions $\mathbf{f}$, primarily
the profile of the uncontrolled TW must be known. In the majority
of cases this profile can be obtained only numerically or experimentally.
Especially in the latter case measurements must be sufficiently accurate
to determine the Goldstone mode $\mathbf{U}_{c}'\left(x\right)$.
Additionally, the propagation velocity $c$ and the invertible coupling
matrix $\mathcal{G}$ are needed. For stationary control Eq. \eqref{eq:FredholmIntegralEquation}
additionally the value of the diffusion coefficient $D$ is required.
Remarkably, the knowledge of the nonlinearity $\mathbf{R}\left(\mathbf{u}\right)$
is not necessary for the calculation of the control functions. This
makes the method powerful for applications where details of the underlying
kinetics $\mathbf{R}\left(\mathbf{u}\right)$ are only approximately
known but the wave profile can be measured with required accuracy.
Examples do not not only include chemical and biological applications
but also population dynamics and spreading diseases \cite{murray1993mathematical}.
Because TW profiles $\mathbf{U}_{c}\left(x\right)$ decay exponentially
fast as $x\rightarrow\pm\infty$, the control Eq. \eqref{eq:SpatiotemporalControl}
is usually localized. If the coupling matrix $\mathcal{G}$ is not
invertible and the RD system is of the form Eq. \eqref{eq:HodgkinHuxleyType},
a control function can still be derived, however, more detailed knowledge
of the reaction kinetics is required, see Eq. \eqref{eq:activatorcontrol}.
In all cases considered the spatio-temporal control Eq. \eqref{eq:SpatiotemporalControl}
was found to be close to an optimal control. We emphasize that in
contrast to our method, computation of an optimal control requires
full knowledge of the reaction kinetics and computationally expensive
algorithms.\\
An important issue is reliability of the proposed controls. Large
control amplitudes $A=c-\dot{\phi}$, Eq. \eqref{eq:SpatiotemporalControl},
sometimes destroy the TW and can lead to the spontaneous generation
of waves, as was also observed in \cite{wolff2003gentle}. We demonstrate
such behavior in the supplemental material, see S7 in \cite{supplement}.
In general, the range of protocol velocities $\dot{\phi}$ achievable
by the proposed control method depends on the reaction kinetics, the
parameter values and higher order derivatives of $\dot{\phi}$. A
necessary condition for the EOM Eq. \eqref{eq:EquationOfMotion} to
be valid is the existence of a spectral gap for the operator $\mathcal{L}$,
Eq. \eqref{eq:LinearOperatorL}. For the Fisher equation, we found
a successful position control despite there is no spectral gap. An
additive control attempting to stop the front leads to a front profile
growing indefinitely to $-\infty$, while a multiplicatively coupled
control accomplishes this task without significantly deforming the
front profile, see S4 and S5 in \cite{supplement}.\\
Generalizing the proposed method to higher spatial dimensions allows
a precise control of shapes of RD patterns. These findings as well
as extensions to conservative nonlinear systems and results regarding
the stability of the control method will be published elsewhere.
\begin{acknowledgments}
We acknowledge support by the DFG via GRK 1558 (J. L.) and SFB 910
(H. E.). 
\end{acknowledgments}
\bibliographystyle{apsrev4-1}
\bibliography{literature}

\end{document}